# Memory Erasure using Time Multiplexed Potentials

Saurav Talukdar, Shreyas Bhaban and Murti V. Salapaka
*University of Minnesota, Minneapolis, USA.*

(Dated: March 1, 2017)

We study the thermodynamics of a Brownian particle under the influence of a time multiplexed harmonic potential of finite width. The memory storage mechanism and the erasure protocol realized by time multiplexed potentials are utilized to experimentally realize erasure with work close to the Landauer's bound. We quantify the work done on the system with respect to the duty-ratio of time multiplexing, which also provides a handle to approach reversible erasures. A Langevin dynamics based simulation model is developed for the proposed memory bit and the erasure protocol, which guides the experimental realization. The study also provides insights into transport at the micro scale.

The Landauer's principle, pioneered by Rolf Landauer in 1961, provides a critical link between information theory and thermodynamics of physical systems [1]. It states that there is no process where the work done to erase one bit of information is less than $k_b T \ln 2$ (Landauer's bound), when the prior probability of the bit being in any of the two states is equal [2]. Here, $k_b$ is the Boltzmann constant and $T$ is the temperature of the heat bath.

Numerous analyses have established the Landauer's bound through different approaches [3–7]. The experimental study of Landauer's bound has only recently become viable, enabled by tools that provide access to processes with energetics in the scale of $k_b T$. A first such study in [8] examined Landauer's bound, by employing optical traps to create a double well potential and realize a single bit memory. Bechhoefer et.al. [9, 10] used an anti-Brownian electrokinetic feedback trap and Hong et.al. [11] used nano magnetic memory bits to analyze the Landauer's bound. Studies in [8–11] demonstrate mechanisms for realizing memory and its erasure with energetics approaching Landauer's bound.

In this article, we study the stochastic energetics of transport realized by time multiplexing of a finite width harmonic potential to effectively realize a bi-stable potential. Here a single laser in an optical tweezer setup, is multiplexed at two locations with varying dwell times to create a series of potentials (symmetric as well as asymmetric bi-stable potentials), that effectuates the erasure process. Moreover, experimental variables to realize *reversible* erasure are identified and utilized for approaching the Landauer's bound. A Langevin framework for a Brownian particle under the influence of a time multiplexed laser is developed and is shown to obey quantitative trends observed in experiments. We use our method of shaping the potential, by changing the dwell time of multiplexing of the laser, to erase one bit of information. The ease of implementation and the high-resolution accounting of energetics enabled by photodiode based measurements are other advantages of the method reported. We resort to Sekimoto's stochastic energetics [12–15] framework to quantify the work done on the system in the erasure process. The underlying principles developed in this article are applicable toward the study of transport achieved by time multiplexing of a single potential, and are not restricted to realizations based on optical traps.

*Experiment and Simulation Model for one-bit Memory:* A Brownian particle in a double well potential is used to model a one-bit memory. The memory is designated the state 'zero' if the particle is in the left well, and the state 'one' if it is in the right well. We realize a Brownian particle in a harmonic potential of finite width experimentally, by using a custom built optical tweezer setup to trap a polystyrene bead ($1\mu m$ in diameter) while suspended in a solution of deionized water (near the focus of a laser beam). The bead experiences a harmonic potential of stiffness $k$ up to a distance $w$ on either side of a locally stable equilibrium point beyond which it does not experience any potential gradient due to the trap (see supplementary material [16] for characterization of $k$ and $w$ for the harmonic potential with finite width). A double well potential with two locally stable equilibrium points located at $L$ and $-L$ is then created by alternately focusing a laser between the two locations (termed as time-multiplexing) at a fast rate (using an Acousto Optical Deflector), as compared to the time scales of the dynamics of the bead. The bead represents the thermodynamic system of interest, the surrounding medium acts as a heat bath, and the trapping laser multiplexed at the two locations is the external system, which is coupled to the thermodynamic system. We define duty-ratio $d$ as the fraction of the time-period the laser spends at the location $-L$. The nature of the effective potential experienced by the Brownian particle can be manipulated by adjusting the duty-ratio. The probability distribution $P_d(x)$ of the position $x$ of the bead in thermal equilibrium with the heat bath at temperature $T$ is determined by the position of the bead measured by the photodiode. The potential energy landscape $U(x,d)$ experienced by the bead for a duty-ratio $d$, is determined by the relationship $P_d(x) = Ce^{-U(x,d)/k_b T}$, where $C$ is a normalization constant; $P_d(x)$ is determined by binning the measured position of the bead. Maintaining a duty-ratio of 0.5 results in near identical parabolic potential wells at $L$ and $-L$ as shown in Fig. 1 while duty-ratio greater than 0.5 leads to asymmetric double well potentials as shown in Fig. 2.

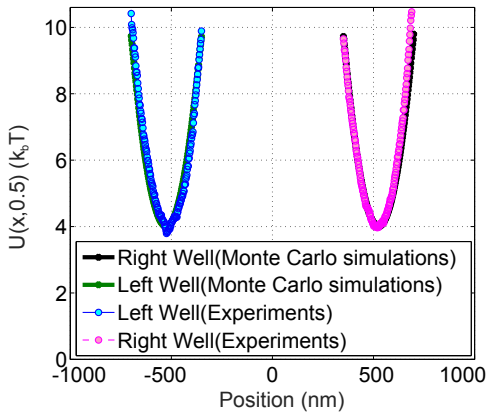

FIG. 1. Double well potential for $L = 550\ nm$ obtained using Monte-Carlo simulations and experiments.

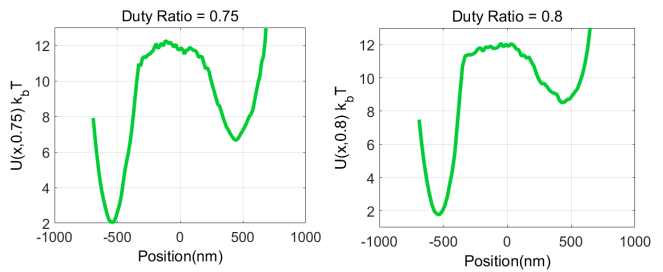

FIG. 2. Effect of duty-ratio on the nature of double well potential. Increasing the duty-ratio at $-L$ from 0.75 to 0.8 increases the asymmetry of the potential.

The bead dynamics is modeled by the overdamped Langevin equation and is given by,

$$-\gamma \frac{dx}{dt} + \xi(t) - \frac{\partial U(x,d)}{\partial x} = 0. \qquad (1)$$

Here, $\gamma$ is the coefficient of viscosity (determined experimentally by step response method [17]) and $\xi(t)$ is modeled as a zero-mean uncorrelated Gaussian noise process with mean, $\langle \xi(t) \rangle = 0$ and correlation, $\langle \xi(t), \xi(t') \rangle = 2\gamma k_b T \delta(t - t')$.

The model for the potential $U(x,d)$ used in (1) incorporates the experimental observation that, for the duration when the laser remains focused at the location $L$ or $-L$, the bead experiences a harmonic potential $\frac{1}{2}k(x - L)^2$ or $\frac{1}{2}k(x + L)^2$ respectively; but beyond the distance $w$ from the locally stable equilibrium points $L$ or $-L$, the bead undergoes a random walk [16, 18]. Here, the stiffness of the laser trap ($k$) and width of the corresponding parabolic potential ($w$) are determined by characterization of the non-ideal harmonic potential obtained due to a single trap, as described in the supplementary material [16]. Monte Carlo simulations performed using (1) and the subsequent potential $U(x,d)$ reconstructed from bead position data, using canonical distribution yield potentials that match closely with experimental observations as seen in Fig. 1. Interestingly, we observe from the Monte Carlo simulations that the stiffness of each of the wells formed at $L$ and $-L$ when the duty-ratio is 0.5 is close to $\frac{k}{2}$, which is half the stiffness $k$ of the trap realized without multiplexing. Moreover, the stiffness of the wells at $L$ and $-L$ obtained from experimental data is also close to $\frac{k}{2}$, thus agreeing with the simulation results.

*Erasure Process:* Erasure is a logically irreversible operation [19], where irrespective of the initial state of the memory, the final state is zero (also known as 'reset-to-zero'operation). A bead in a symmetric double well potential can occupy either well with equal probability; and is used to model a single bit memory. Initially, it is equally likely that the memory occupies the state zero or one, and at the end of erasure process the memory state must be *zero* (the bead must be in the left well). Thus, there is no change in average energy of the bead in an erasure process while the decrease in entropy associated with erasure is $k_b \ln 2$; thereby requiring at least $k_b T \ln 2$ amount of work to be done on the system [2]. The Landauer's bound is applicable to the average work done on the system over many realizations of the bead trajectory; indeed, it is possible to obtain individual erasure realizations where the work done on the system is less than $k_b T \ln 2$. As we demonstrate later, for certain trajectories the work done on the bead is lower than $k_b T \ln 2$ (see Fig. 6).

The Landauer's bound of $k_b T \ln 2$ holds if the erasure process is always successful. Partially successful erasures with success proportion $p$ require at least $k_b T (\ln 2 + p \ln p + (1 - p) \ln(1 - p))$ amount of work to be done on the system [2]. It is important to note that the bound decays rapidly as $p$ decreases from 1; with the bound being $k_b T \ln 2$ for $p = 1$ and zero for $p = 0.5$. In our study, we ensure that $p > 0.95$. For computing the work done on the system, only those trajectories that lead to a successful erasure are considered.

To realize erasure, the duty-ratio of laser time-multiplexing is manipulated according to the following protocol: initially, a symmetric double well potential is realized by employing a duty-ratio of 0.5. Then, for raising the well at $L$ and lowering the well at $-L$ (asymmetric double well potential), the duty-ratio at $-L$ is increased to values greater than 0.5 for a time duration $\tau$. As a result, over the time duration $\tau$, the laser spends more time focused at $-L$ than at $L$, enabling an asymmetric potential landscape as is observed in Fig. 2. Increasing the duty-ratio results in a lower barrier height for the right to left transition than for the left to right transition. It thus favors the transport of the bead from the right to the left well, if it was initially in the right well as shown in Fig. 3 (a) and retains the bead in the left well if it was initially in the left well as shown in Fig. 3(b). Finally, we revert the duty-ratio to 0.5. For the erasure duration $\tau$ chosen in this article, that is the duration for which we have an asymmetric double well potential (see [16]), the success proportion $p$ obtained from simulations is shown in Fig. 4. It is seen that the duty-ratio of 0.65 yields success proportion significantly less than 0.95, while a duty-ratio $> 0.7$ shows a success proportion greater than 0.95. Similar trends are reflected from experiments as seen in Fig. 4.

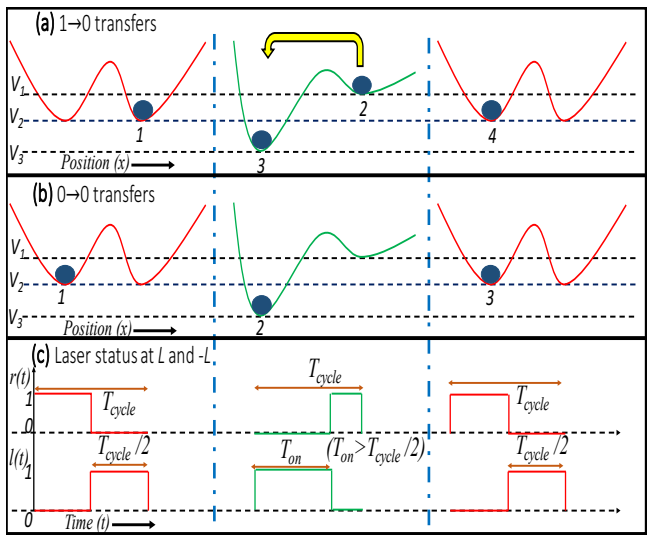

FIG. 3. (a) Schematic showing erasure process, with bead initially in the right well. The initial bead position is 1 (right well), with potential energy $V_2$ The duty-ratio $d$ at left well is then increased, which lifts the bead and takes it to position 2 with energy $V_1$. Thermal fluctuations enables the bead to cross the barrier and reach position 3, with energy $V_3$. Decreasing the duty-ratio back to 0.5 lifts the bead to position 4, which has energy $V_2$. The process $1 \to 2 \to 3 \to 4$ is the erasure process. (b) Schematic showing erasure process, with bead initially in the left well. Here, the process $1 \to 2 \to 3$ is the erasure process. (c) The signals $r(t)$ and $l(t)$ denote the presence/ absence status of the laser at $L$ and $-L$ respectively. A value of 1 means present and 0 means absent. To ensure a duty-ratio greater than 0.5, we maintain $d = T_{on}/T_{\text{cycle}} > 0.5$.

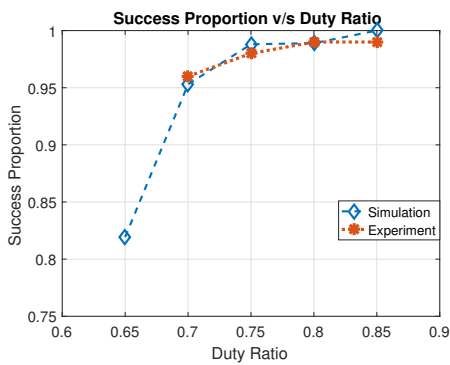

FIG. 4. Effect of duty-ratio on success proportion $p$. duty-ratio of 0.65 has a success proportion of 0.82.

*Erasure Thermodynamics:* We now utilize the stochastic-energetics framework for Langevin systems [14, 15] and quantify the work done on the system, associated with erasure process realized by manipulation of duty-ratios. The external system does work on the bead by changing the duty-ratio, which results in modifying the potential felt by the bead. For an erasure process, the work done on the bead $dW$ is given by,

$$dW = \sum_j [U(x(t_j), d(t_j^+)) - U(x(t_j), d(t_j^-))], \quad (2)$$

where $d$ denotes the discontinuous parameter (here, the duty-ratio) changed by the external system, and $t_j$ denotes the time instances when the parameter was changed ($t_j^-$ and $t_j^+$ denote the instants just before and after changing the parameter respectively).

Landauer's bound can be reached when the erasure process is performed in a quasi static manner, with success proportion of erasure being one. For an erasure performed over a large but finite duration $\tau$, the average work done on the system is [13],

$$\langle dW \rangle = dW_{Landauer} + \frac{B}{\tau} \quad (3)$$

where, $dW_{Landauer} = k_b T \ln 2 = 0.693 k_b T$. The duration for which an asymmetric double well potential is realized, $\tau$, is a scalar multiple of the exit time of the bead from the right well, denoted by $\tau_e$. It is known that $\tau_e \propto \frac{\exp(\delta U_r)}{\sqrt{k_r}}$, [20] where $\delta U_r$ is the barrier height of the right well, $k_r$ is the stiffness of the right well and $\exp(.)$ is the exponential function. Note that $d-0.5$ is indicative of the asymmetric nature of the double well potential; higher the value more is the asymmetry. We determine the dependency of $\delta U_r$ and $k_r$ on the $\frac{1}{d-0.5}$ (inverse of the deviation of duty ratio from 0.5) empirically, to obtain $\tau \propto \tau_e \propto \frac{\exp(\frac{1}{d-0.5})}{\sqrt{\frac{1}{d-0.5}}}$ (see supplementary material for details), and substituting it in (3) for $\tau$ leads to,

$$\langle dW \rangle = dW_{Landauer} + B \frac{\exp(-\frac{1}{d-0.5})}{\sqrt{d-0.5}}. \quad (4)$$

In the limit of $d \downarrow 0.5$, the asymmetry of the double well potential vanishes, and the time duration $\tau$ required for successful erasures (with $p = 1$) is large when compared to time scales of bead dynamics, thereby resembling a quasi-static process. Thus, the duty-ratio provides a handle to realize quasi-static erasure processes using time multiplexed potentials.

The average work done on the bead $\langle dW \rangle$, for duty-ratio $d > 0.7$ obtained through simulations and experiments is shown in Fig. 5(a) and Fig. 5(b) respectively; where the trends are similar. It is seen that as the duty-ratio is reduced, the average work done on the bead decreases. For a duty-ratio of 0.7, average work done on the system obtained through simulations is $0.71 \pm 0.035 k_b T$. Experimentally for duty-ratio of 0.7, the average work done on the bead is determined to be $0.9 \pm 0.106 k_b T$. The average work of $0.9 \pm 0.106 k_b T$ to erase a bit of information is the closest to the Landauer's bound of $k_b T \ln 2$ reported. The primary sources of error in the average work done on the system computed from position measurements are introduced by the photodiode based measurement scheme, which is estimated to be in the order of $10^{-3} k_b T$ (see supplementary material [16], [21]).

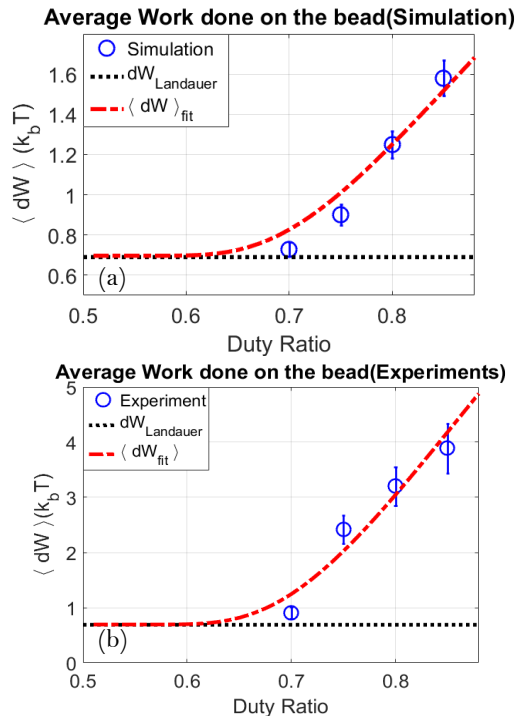

FIG. 5. The blue circles represent the average work done on the bead obtained using simulations in (a) and experiments in (b) for duty-ratio of $0.7, 0.75, 0.8, 0.85$. The vertical lines represent the standard error in mean for each duty-ratio. The black dotted line denotes the the Landauer bound of $k_bT\ln 2$. The red dotted line is the fit with the free parameters $A$ and $B$.

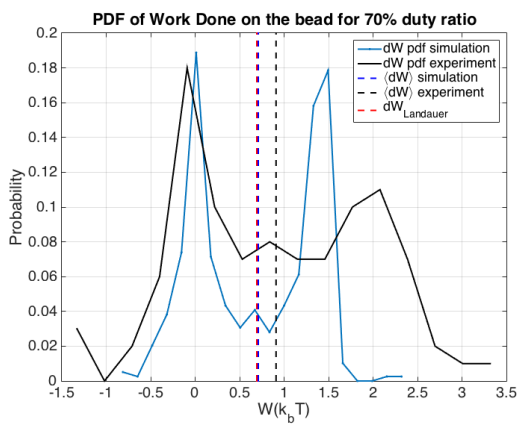

FIG. 6. Distribution of work done on the bead obtained from simulations and experiments for a duty-ratio of 0.7. The nature of the distribution is bimodal.

We fit the model derived in (4), $\langle dW \rangle_{fit} = A + B\frac{\exp(-\frac{1}{d-0.5})}{\sqrt{d-0.5}}$ to the average work done on the bead obtained by simulations and experiments for various duty-ratio values, with $A$ and $B$ being free parameters (see Fig. 5). Using simulation data we obtain $A = 0.697 k_bT, B = 8.25 k_bT$, whereas for experimental data we have $A = 0.701 k_bT, B = 35.04 k_bT$. It is seen that $A$ (which represents the average work done on the system in the quasi-static case) has a value close to the Landauer bound of $k_bT\ln 2 \,(= 0.693 k_bT)$, in both simulations as well as experiments.

The distribution of work done while erasing a bit at a duty-ratio of 0.7, obtained from simulations and experiments is shown in Fig. 6. It is evident that for a fraction of trajectories, the work done on the bead is less than the Landauer's bound; indeed, for some trajectories it is negative. However, the mean of the distribution is close to the Landauer's bound. Moreover, a bimodal nature of the distribution is evident. The mode on the right corresponds to work done for the right to left (or $1 \to 0$) transfers and mode on the left corresponds to left to left (or $0 \to 0$) transfers. The characteristics of the simulation data are confirmed by experiments as shown in Fig. 6.

In summary, the energetics of a Brownian particle influenced by a time-multiplexed potential was analyzed using Sekimoto's framework. The duty-ratio of how the laser is multiplexed is used to realize erasure of one bit of information. The deviation of the external parameter of duty-ratio from 0.5 provides an effective way of controlling the approach of an erasure process toward a quasi-static process. Our analysis indicates that as the duty-ratio is brought closer to 0.5, the average work done on the system approaches the Landauer's bound of $k_bT\ln 2$.

*Conclusions:* In this article, we study the thermodynamics of a Brownian particle influenced by the time multiplexing of a single non-ideal harmonic potential. A Monte Carlo simulation framework for a Brownian particle under the influence of a time multiplexed laser is also developed and is shown to obey quantitative trends observed in experiments. We demonstrate that the duty-ratio provides a handle on the speed of the erasure process and its approach to reversibility. It is established through experiments and simulations that realizing erasure with the duty-ratio nearing 0.5 results in the average work done approaching $k_bT\ln 2$; which is the minimum average work required to erase one bit of information. Furthermore, the method is easy to implement on a standard optical tweezer setup. The insights obtained from this article can be potentially leveraged to realize practical devices that yield erasure mechanisms with energetics in the order of $k_bT\ln 2$.

*Acknowledgement:* We would like to thank Arun Majumdar, Standford University for initial discussion about the problem, Srinivasa Salapaka, University of Illinois, Urbana Champaign for his comments, Tanuj Aggarwal, Cymer LLC and Subhrajit Roychowdhury, GE Research for their contribution towards building the optical tweezer setup. The research reported is supported by National Science Foundation under the grant CMMI-1462862.

*Author Contributions:* S.T. and S.B contributed equally to this work. S.T., S.B. and M.S. conceptualized the work. S.T. and S.B. developed the computational framework and conducted experiments.